# Cybersecurity policy adoption in South Africa: Does public trust matter?

**Mbali Nkosi**, u22500172@tuks.co.za

**Mike Nkongolo**, mike.wankongolo@up.ac.za

**Department of Informatics, University of Pretoria**

## Abstract

*Public trust, or the lack thereof, can be a hindrance to the development and adoption of cybersecurity policies. This study examines how public perception influences the implementation and adoption of cybersecurity frameworks in South Africa. Using the PRISMA methodology, a systematic literature review was conducted across reputable scholarly databases, yielding 34 relevant sources aligned with predefined inclusion criteria. Cybersecurity, governance, trust, privacy, cybercrime, and public opinion emerged as dominant thematic clusters. Bibliometric and thematic analyses, supported by network visualisations, revealed that while trust and public sentiment significantly affect cybersecurity policy adoption globally, these factors exert minimal influence within the South African policy landscape, despite the country's high cybercrime prevalence. In response, the study proposes a trust-centric policymaking framework designed to integrate public perception as a proactive dimension of cybersecurity governance. This framework aims to prevent trust deficits from becoming barriers to policy effectiveness in South Africa and offers guidance for restoring trust in contexts where it has eroded. The paper concludes with key limitations and recommendations for future research.*

**Keywords:** Cybersecurity; Public trust; Government; Public perception; Cybersecurity frameworks

## 1. INTRODUCTION

Public perception of government influences policy implementation (Okunoye, 2022). Studies show that trust in government affects the adoption and success of cybersecurity measures; low trust can limit public compliance and reduce the effectiveness of these efforts (Ball, Esoposti, Dib, Pavione & Santiago-Gomez, 2018; Djuric, 2024). Governments must therefore work to build confidence and address mistrust regarding security mechanisms (Okunoye, 2022). The growth of E-Government services has increased the need for comprehensive data protection to safeguard citizens' information from breaches and cyber threats (Hossain, Yigitcanlar, Nguyen & Xu, 2024).

In South Africa, Pieterse (2021) reported 74 cyber incidents between 2010 and 2020, including 27 attacks on public sector platforms due to inadequate cybersecurity resources and increasing IT-related threats. These events highlight weaknesses in national policies such as the National Cybersecurity Policy Framework (NCPF) (Sutherland, 2017; Verkijika & De Wet, 2018). Political trust is a key factor in policy execution; lack of trust can delay or hinder cybersecurity initiatives, as seen in Nigeria and Australia (Bal et al., 2018; Hilowle, Yeoh, Grobler, Pye & Jiang, 2022b; Okunoye, 2022). Governments often need to engage with citizens to rebuild trust and ensure effective adoption of policies (Okunoye, 2022). Public perception also shapes policymaking and cybersecurity framework development, requiring strategies that balance local needs with international standards (Joubert & Wa Nkongolo, 2025; Craig, Johnson & Gallop, 2023). Despite its importance, there is limited research on how public trust influences cybersecurity policy adoption. This study addresses this gap through a systematic literature review, aiming to develop a framework that integrates public perception into cybersecurity policy implementation in South Africa. The main research question is: *How does public trust in government influence the adoption and implementation of national cybersecurity policies in South Africa?* The paper is organized as follows: Section 2 provides background and literature context; Section 3 details methodology and data analysis; Section 4 presents visual findings; Section 5 offers comprehensive analysis; Section 6 consolidates results; and Section 7 identifies research gaps.

## 2. LITERATURE BACKGROUND

Cybersecurity has become crucial for Africa, particularly South Africa, due to rapid economic growth and widespread technology adoption (Barros & van Niekerk, 2023). Understanding the role of public trust in implementing cybersecurity frameworks requires examining Africa's cybersecurity landscape, challenges, and policy acceptance.

### Cybersecurity, Trust, and Policy Acceptance

Cybersecurity involves mechanisms to protect information from theft, vulnerabilities, and unauthorized access through frameworks, policies, and processes (Delgado, Esenarro, Regalado & Reátegui, 2021). Public policy acceptance reflects how communities perceive new policies as favorable or unfavorable (Rodriguez-Sanchez et al., 2018). Trust is defined as public optimism in government sincerity, distinct from "trustworthiness" (Okunoye, 2022). Political trust in cybersecurity thus refers to citizens' confidence in security measures and the institutions implementing them (Alam, Ahmed, Dahli & Alam, 2024; Delgado et al., 2021; Hossain et al., 2024; Okunoye, 2022).

**Cybercrime in Africa**

Rising ICT use has increased cyberthreats and cybercrime across Africa (Barros & van Niekerk, 2023; Kshetri, 2019; Maphosa, 2024). Major incidents include 2017 losses in South Africa, Nigeria, and Kenya ($157M, $694M, and $210M respectively) and Ethiopia's cyberattacks on the Grand Ethiopian Renaissance Dam (Keleba, Tabona & Maupong, 2022). Mozambique experienced over 30 government website attacks in 2022, though without data loss (Barros & van Niekerk, 2023).

**Cybersecurity Efforts and Challenges**

African governments, including South Africa, Nigeria, Zimbabwe, Mozambique, and Tanzania, have enacted cybersecurity policies and frameworks (Barros & van Niekerk, 2023; Kabanda, 2020; Maphosa, 2024; Pieterse, 2021; Sutherland, 2017). These include data protection acts and cybersecurity frameworks (Olarindem Yebisi, Anwana, Ogundele & Awodiran, 2024b; Olukoya, 2022). However, implementation faces challenges: limited public awareness, low prioritization by governments, miscoordination of resources, potential surveillance, censorship, and reliance on insufficient training (Barros & van Niekerk, 2023; Maphosa, 2024; Mganyizi, 2023; Mwogosi & Simba, 2025; Pieterse, 2021; Sutherland, 2017). Public opinion and trust have influenced policy outcomes but were rarely integrated into policy decisions (Maphosa, 2024; Mganyizi, 2023; Mwogosi & Simba, 2025). This gap highlights the need to examine how trust affects the adoption and effectiveness of cybersecurity measures. This study uses the PRISMA model for a systematic literature review to investigate the state of cybersecurity in South Africa and how trust, mistrust, or distrust in government affects policy adoption. The goal is to develop a framework applicable in the South African context.

## 3. RESEARCH METHOD

A systematic literature review (SLR) synthesises secondary qualitative data from academic journals (Snyder, 2019) examining the impact of public trust on the government's ability to adopt and implement cybersecurity measures successfully. The findings address the research question and evaluate the study's hypothesis, contributing to the development of a framework to help government institutions foster trust and support robust cybersecurity implementation. This study employs the PRISMA model to ensure transparent data collection and reporting (Trifu, Smîdu, Badea, Bulboacă & Haralamibe, 2022). Articles were selected using defined criteria from trusted databases, including SciELO and Taylor & Francis, yielding 34 relevant papers. Key information was extracted through bibliometric and thematic methods to provide a clear and transparent view of the review process (Büchter, Romney, Mathes, Khalil, Lunny, Pollock, Puljak, Tricco & Pieper, 2023). Thematic analysis (Braun & Clarke, 2006) explored relationships between identified themes, while bibliometric analysis examined trends such as study location (Mukhtar, Shad, Ali Woon & Waqas, 2024). Main themes identified such as "Cybersecurity," "Government," "Trust," "Privacy," "Cybercrime," and "Public Opinion" guided the discussion of findings and their relevance. The study evaluates the role of institutional trust in policymaking using Institutional Trust Theory (Godefroit, Langer & Meuleman, 2017) alongside the cyber-policy framework by Joubert and Wa Nkongolo (2025), which emphasizes multi-stakeholder involvement. Based on this, the study tests the following hypotheses: *the null hypothesis (H0) posits that negative public perceptions of government structures hinder the adoption and implementation of cybersecurity frameworks in South Africa, whereas the alternative hypothesis (H1) proposes that integrating public trust, technical expertise, and multi-stakeholder involvement enhances the effectiveness, sustainability, and public acceptance of cybersecurity policy implementation*.

## 3.1 DATA SOURCES AND SEARCH TERMS

To obtain academic journals, search strings composed of keywords relevant to the topic were used to query academic databases (Tranfield, Denyer & Smart, 2003; Trifu et al., 2022; Asubiaro, Onaolap & Mills, 2024; Raashida, Arisha, Sumeera & Wani, 2021). Table 1 presents the search strings used across different databases and the rationale for each.

**Table 1: Search terms and data sources**

| Search String | Reason | Database Searched | Number of Sources |
|---|---|---|---|
| ("government institutions" OR "state institutions" OR "government" OR "state" OR "federal institutions" OR | This search string yielded sources that discuss trust in relation to government institutions in | Web of Science - All Databases, | 3395 |
| "federal" OR "government agencies" OR "agencies" OR “local government” ) AND ("cybersecurity framework" OR "cybersecurity policy" OR "cybersecurity implementation" OR "cybersecurity adoption" OR "cybersecurity" OR "data security" OR "information security" OR "information technology security" OR “national security” ) AND ("public trust" OR "citizen trust" OR "trust" OR "public confidence" OR "mistrust" OR "distrust" | relation to cybersecurity. This will assist in the investigation of the positive and negative implications of public trust on cybersecurity adoption and implementation. | Scopus | |
| ("cybersecurity framework" OR "cybersecurity policy" OR "cybersecurity implementation" OR "cybersecurity adoption" OR "cybersecurity" OR "data security" OR "information security" OR "information technology security" OR “national security” ) AND ("South Africa" OR "South African") | This search string yielded sources that discuss cybersecurity policies frameworks or their implementation. This will assist in gathering papers that discuss the current state of cybersecurity in South Africa, including any shortcomings that may exist in which the development of a framework would be useful. | Web of Science - All Databases, Scopus | 517 |
| ("government institutions" OR “local government” ) AND ("cybersecurity framework" OR "data security" OR "information security") AND ("public trust" OR "citizen trust" OR "individual trust") | Due to boolean operator and character limit constraints, the search string had to be shortened when utilising the ScienceDirect database | ScienceDirect | 1278 |
| ("cybersecurity framework" OR "cybersecurity implementation" OR "cybersecurity adoption" OR "cybersecurity" OR "data security" OR "information security" OR “national security” ) AND ("South Africa" OR "South African") | Due to boolean operator and character limit constraints, the search string had to be shortened when utilising the ScienceDirect database | ScienceDirect | 2334 |

## 3.2 SELECTION CRITERIA

The criteria consisted of inclusion and exclusion rules, applied respectively to each paper retrieved from the databases using the search terms listed in Table 1 (Tranfield, Denyer & Smart, 2003; Tranfield *et al.*, 2003; Trifu *et al.*, 2022). The inclusion criteria includes journal articles that discuss public trust, mistrust or distrust and the implications of that on successful cybersecurity implementation. Or how building trust has been effective in implementing cybersecurity nationally and the discussion of the state of cybersecurity in South Africa. The exclusion criteria include journal articles not written in English, articles that are not peer-reviewed, sources that are not journal articles or conference papers, duplicate entries, papers without titles, and papers whose titles do not contribute to the topic under investigation.

### 3.3 PRISMA FLOWCHART

The PRISMA flowchart shows four steps depicting the execution of the inclusion and exclusion criteria (Trifu *et al.*, 2022) as stipulated in Sections 3.2.1 and 3.2.2 (Figure 1).

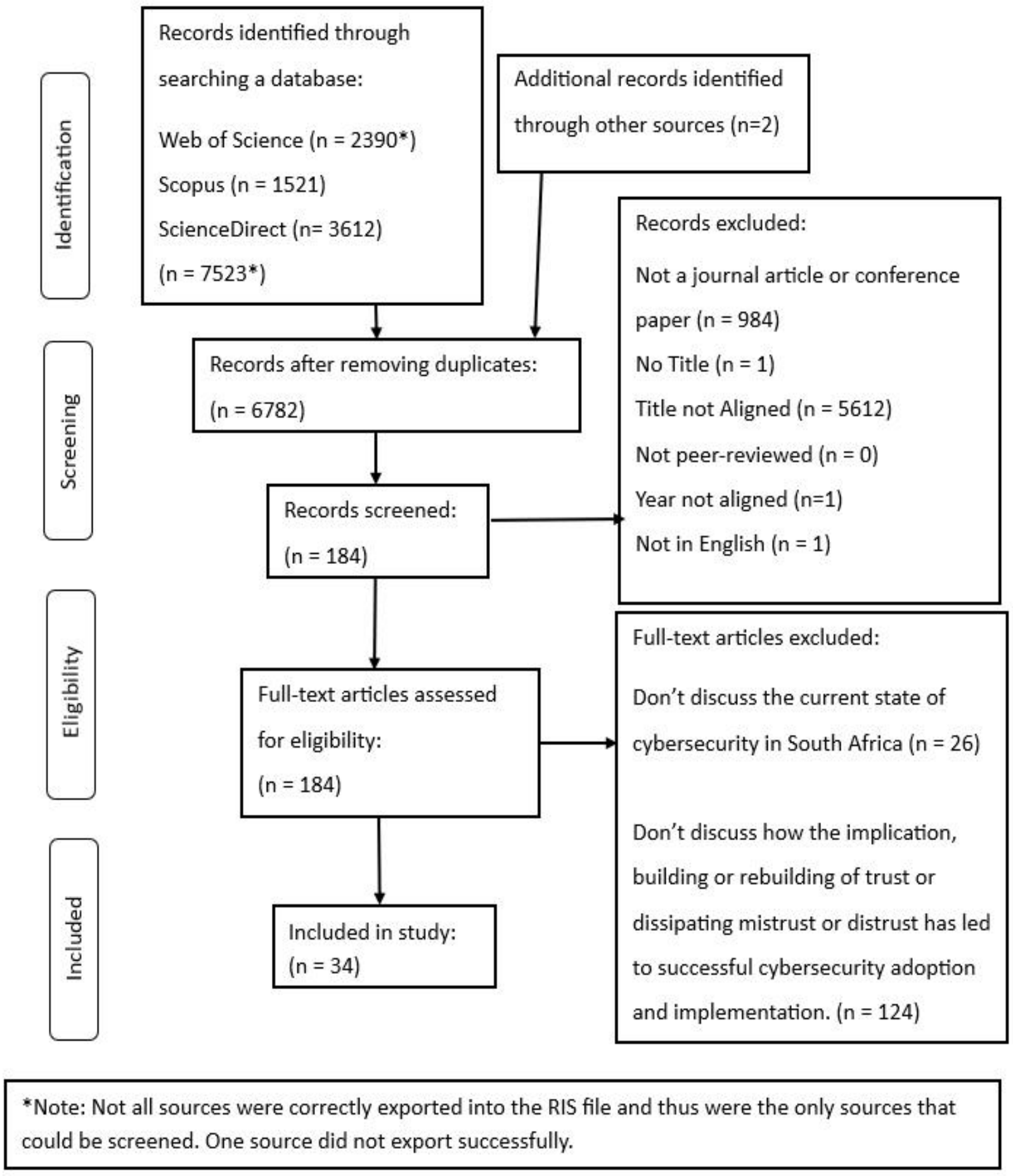

**Figure 1: PRISMA flowchart**

The flowchart (Figure 1) illustrates each step of the selection process, culminating in the identification of sources eligible for inclusion and data analysis. Initially, 7,523 sources were retrieved from Web of Science, Scopus, and ScienceDirect, with two additional sources from SciELO and Taylor & Francis. During screening, 7,339 sources were removed due to duplicates, language, incorrect source type, non-contributory or missing titles, and date restrictions, leaving 184 sources. A full-text eligibility assessment excluded 150 sources that did not address South Africa's cybersecurity context or the role of trust. Ultimately, 34 sources met all criteria and were included in the final review (Figure 1).

### 3.4 DATA EXTRACTION

The papers included in Section 3.3 were collected, and relevant data were extracted using the chosen methods to objectively review and synthesise the findings (Tranfield, Denyer & Smart, 2003; Büchter et al., 2023). Data extraction employed both bibliometric and thematic approaches.

### 3.5 DATA ANALYSIS

The data extracted in Section 3.4 were integrated and summarised to facilitate discussion of findings and generate insights (Tranfield, Denyer & Smart, 2003). Thematic analysis was used to identify, analyse, and report patterns from the extracted data (Braun & Clarke, 2006), while bibliometric analysis provided deeper insights into the included publications (Mukhtar, Shad, Ali Woon & Waqas, 2024). As the study's primary focus is cybersecurity, the term "Cybersecurity" is used to encompass "information security" and "data security," with all discussions framed in the context of cybersecurity adoption.

**Bibliometric Analysis**

Figure 2 shows that most papers relevant to this research were published between 2020 and 2024, with a notable peak in 2017, indicating that the data reflect current cybersecurity topics (Figure 3).

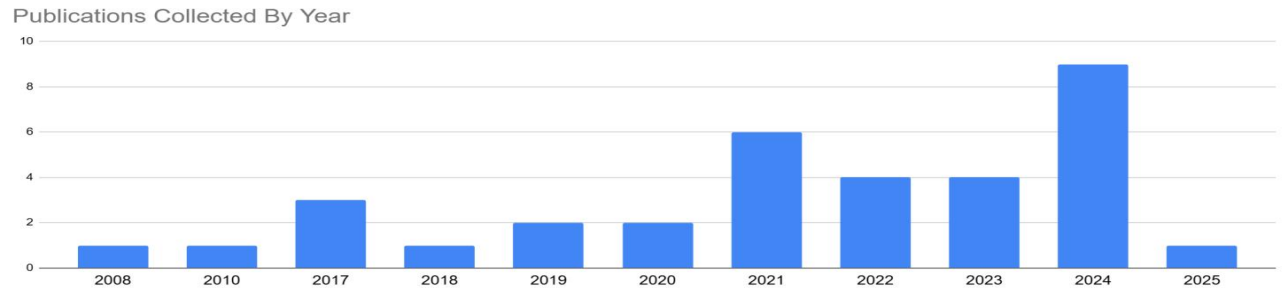

**Figure 2: Number of publications by year**

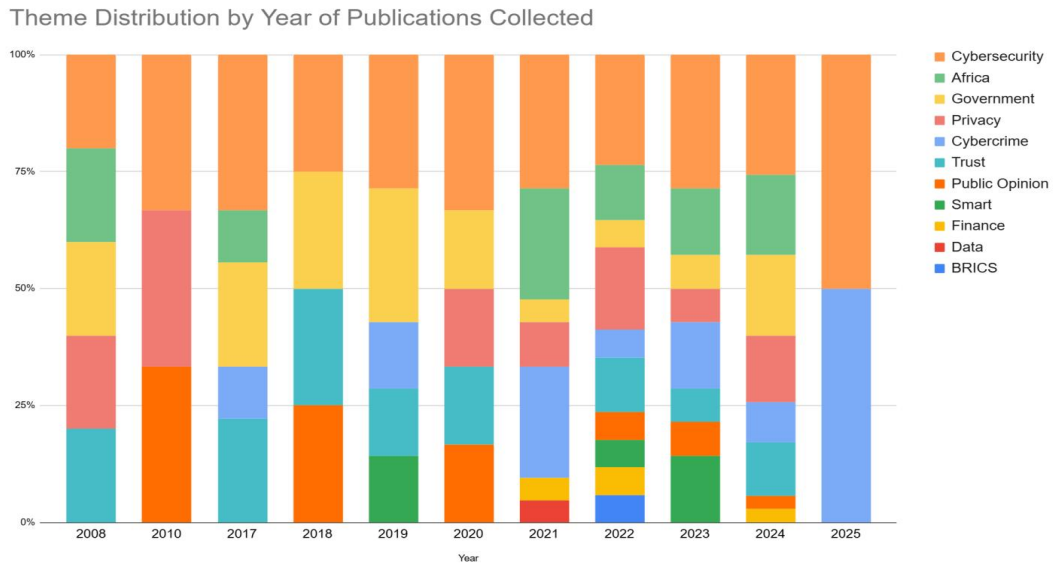

**Figure 3: Theme distribution by year of study**

Earlier publications remain valuable for identifying past trends and assessing the evolution of cybersecurity globally and in South Africa. Figures 3 to 5 further show that the main themes of the study were discussed consistently across the years, supporting the longitudinal assessment of theme development.

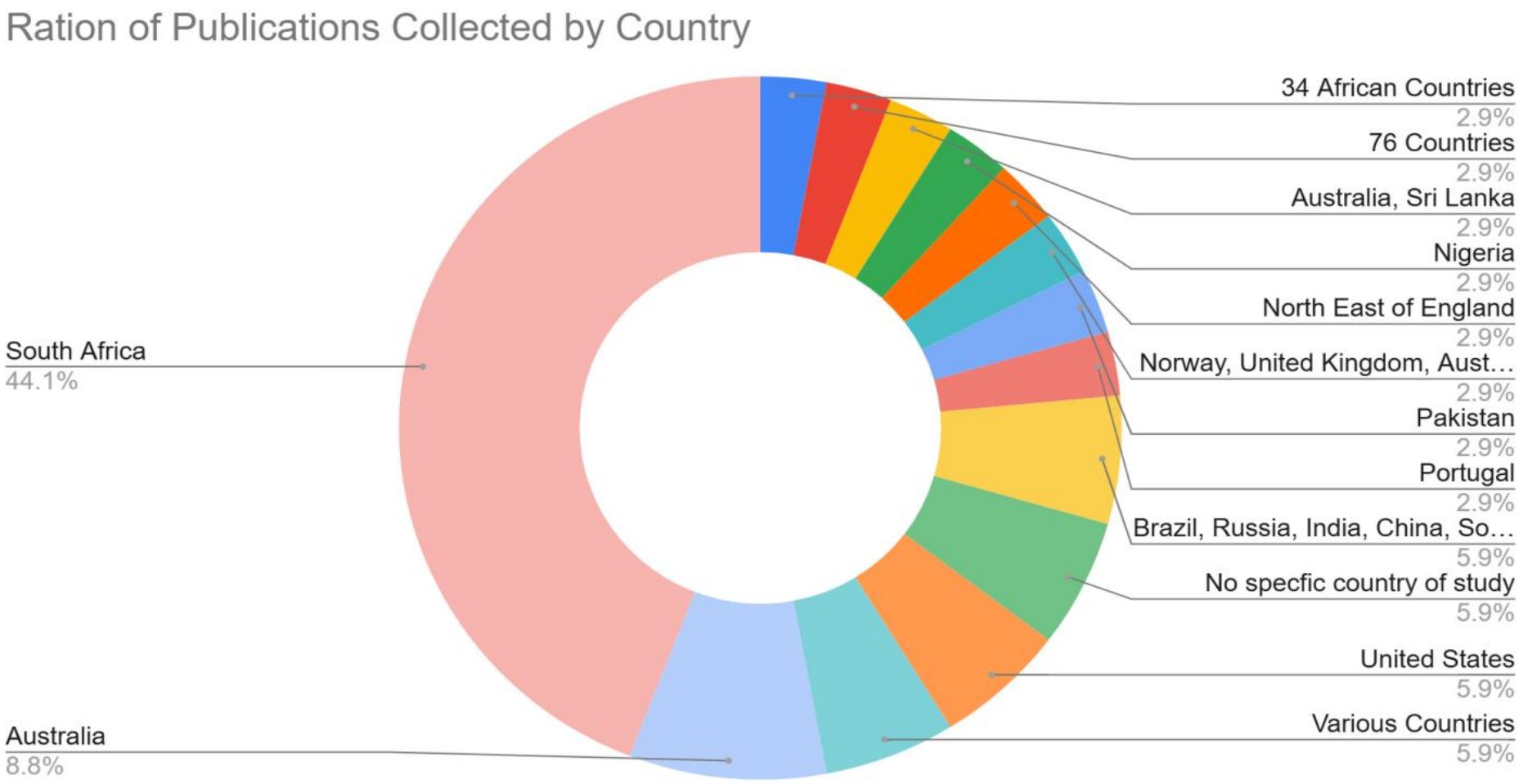

**Figure 4: Number of publications by country of study**

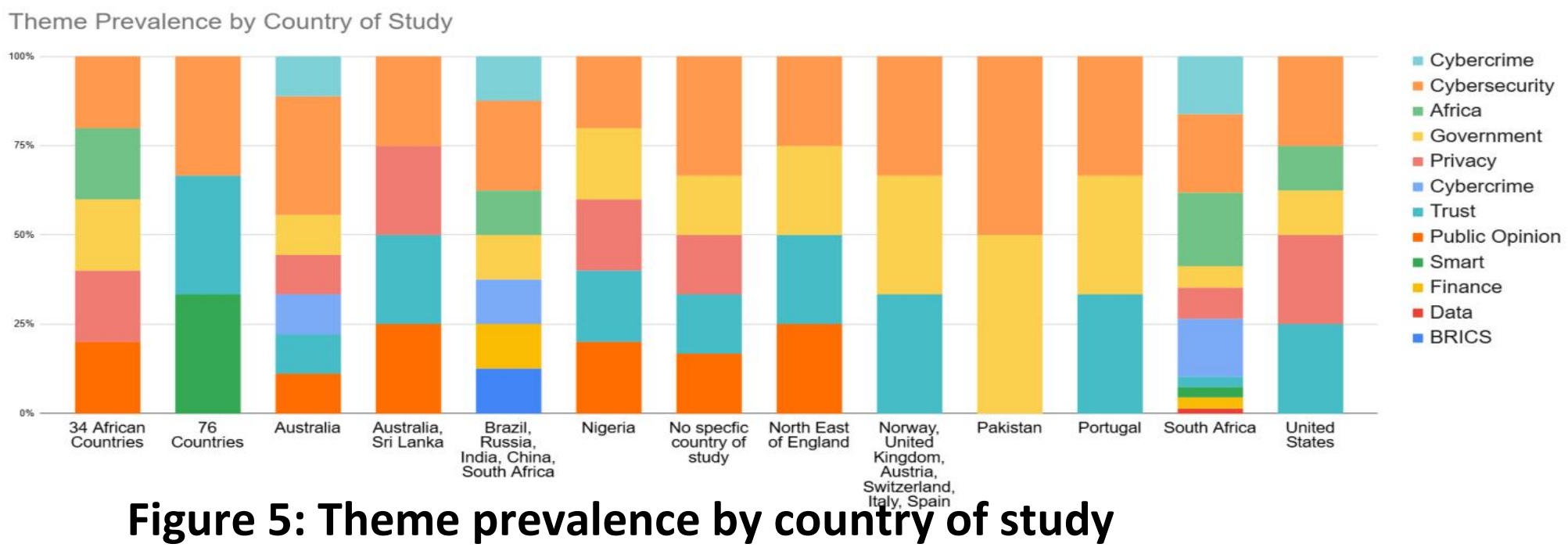

**Figure 5: Theme prevalence by country of study**

Figure 4 illustrates the number of studies conducted across different countries, with South Africa accounting for 44.1% of the publications, providing a detailed view of the current cybersecurity landscape in the country. The presence of studies from other nations enables future comparative research on cybersecurity contexts. This distribution reflects the search terms used. Figure 5 shows the prevalence of themes across these countries. In South Africa, studies on cybersecurity rarely address trust or public opinion, whereas these themes are more common in research from Australia and broader African contexts, including multi-country analyses. This indicates that trust and public perception may not yet be a priority in South African cybersecurity policymaking.

**Thematic Analysis**

The data corresponding to the identified themes were analysed and visualised in Figures 6–8 to provide insights from the extracted data. Figure 6 shows that "Cybersecurity" is the most prevalent theme, with 47 occurrences across the 34 sources, peaking in 2021 and 2024. While some studies addressed subthemes nested under Cybersecurity, the majority discussed cybersecurity holistically: 14 sources focused solely on cybersecurity, while 20 included one or more subthemes. For this study, the themes "Finance," "Smart," "Data," "Europe," "Africa," and "BRICS" (Figures 5-6) were excluded due to limited data and non-contributory relevance.

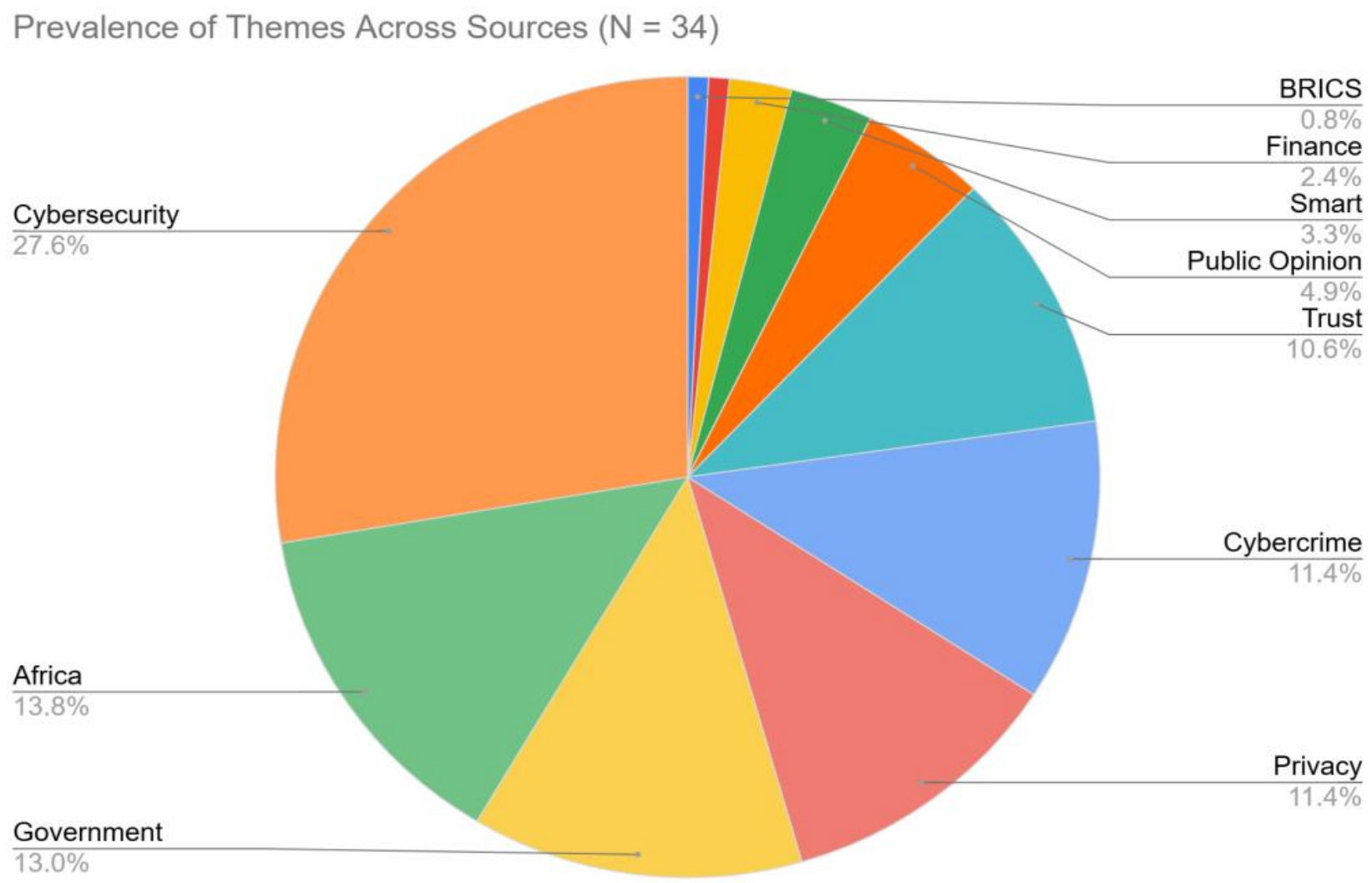

**Figure 6: Prevalence of themes across sources**

## 4. FINDINGS

Figures 7 shows the links and relationships that are prominent.

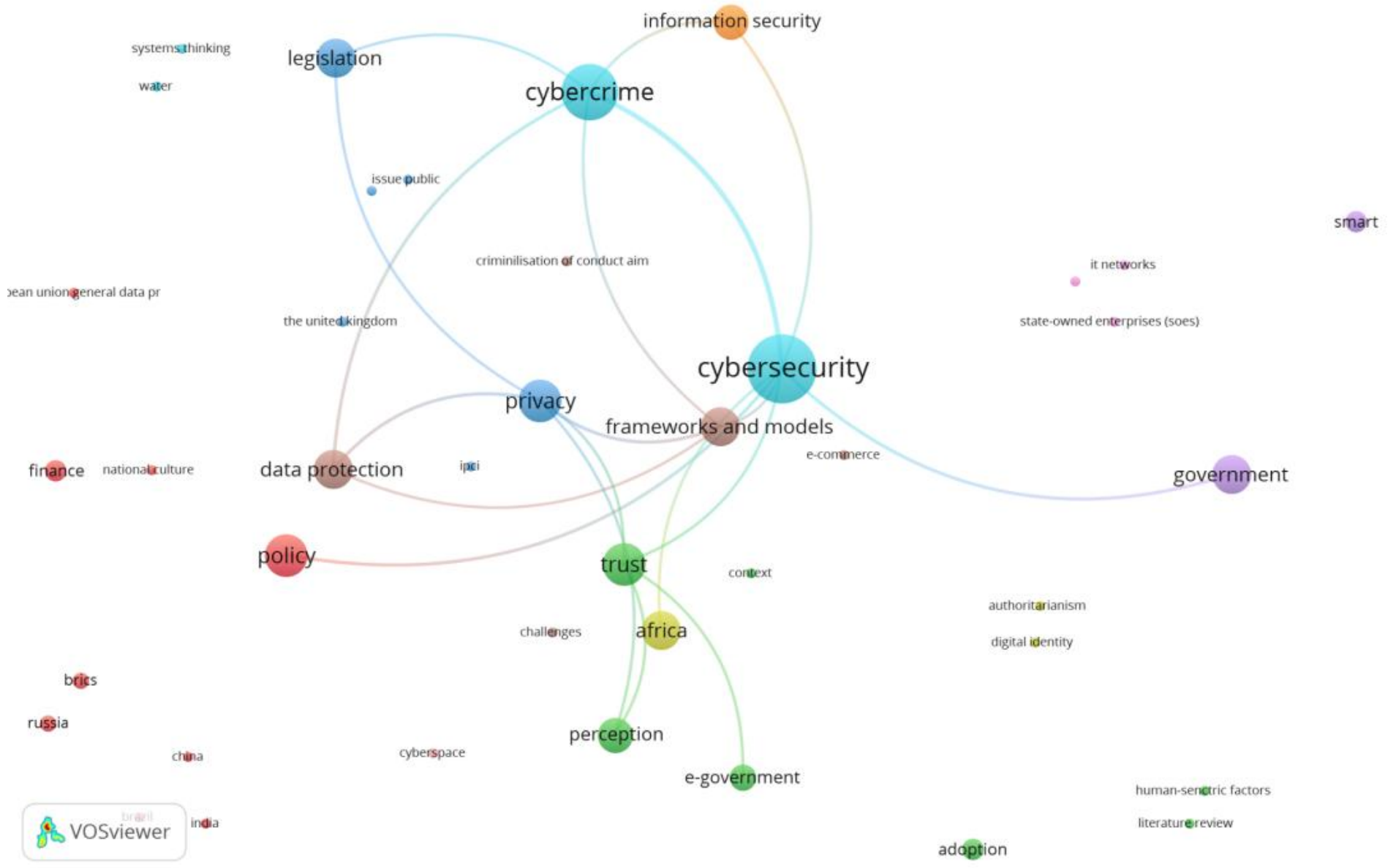

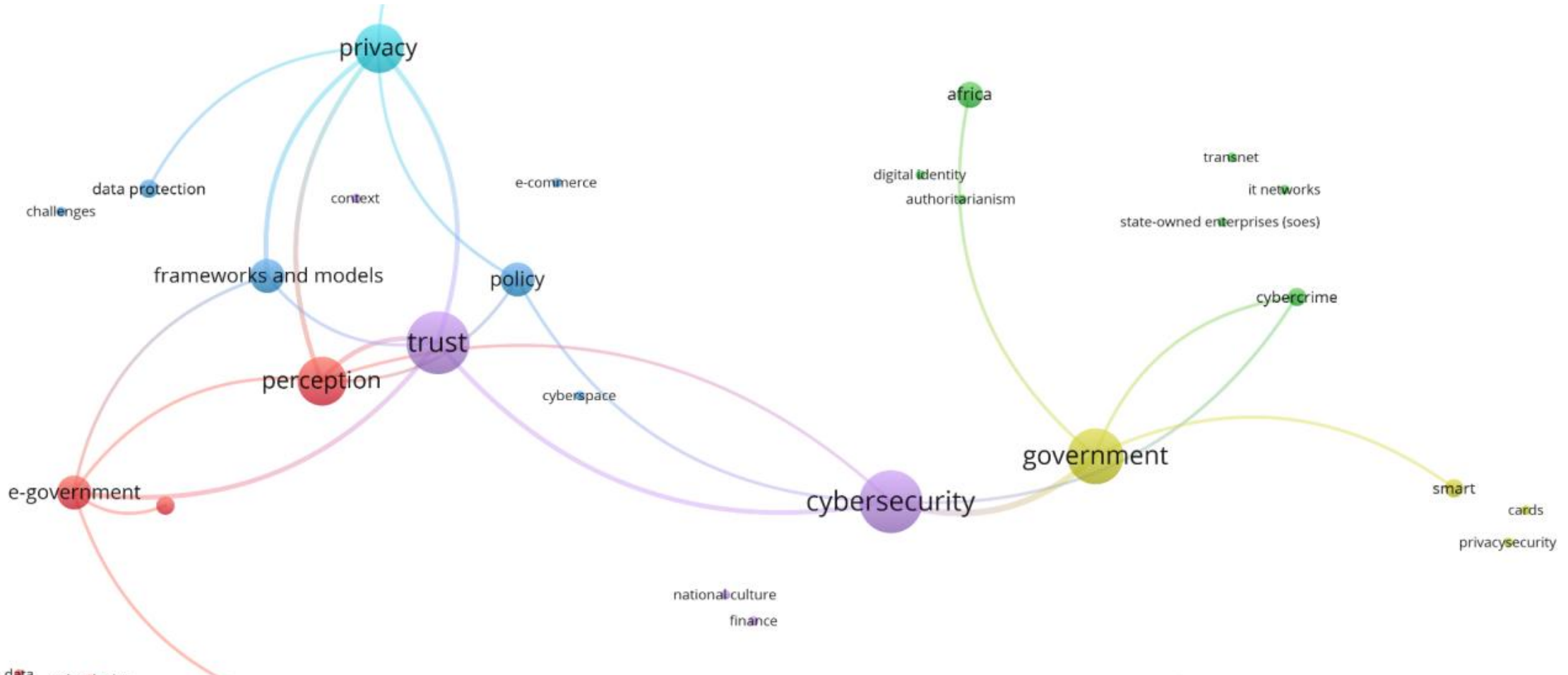

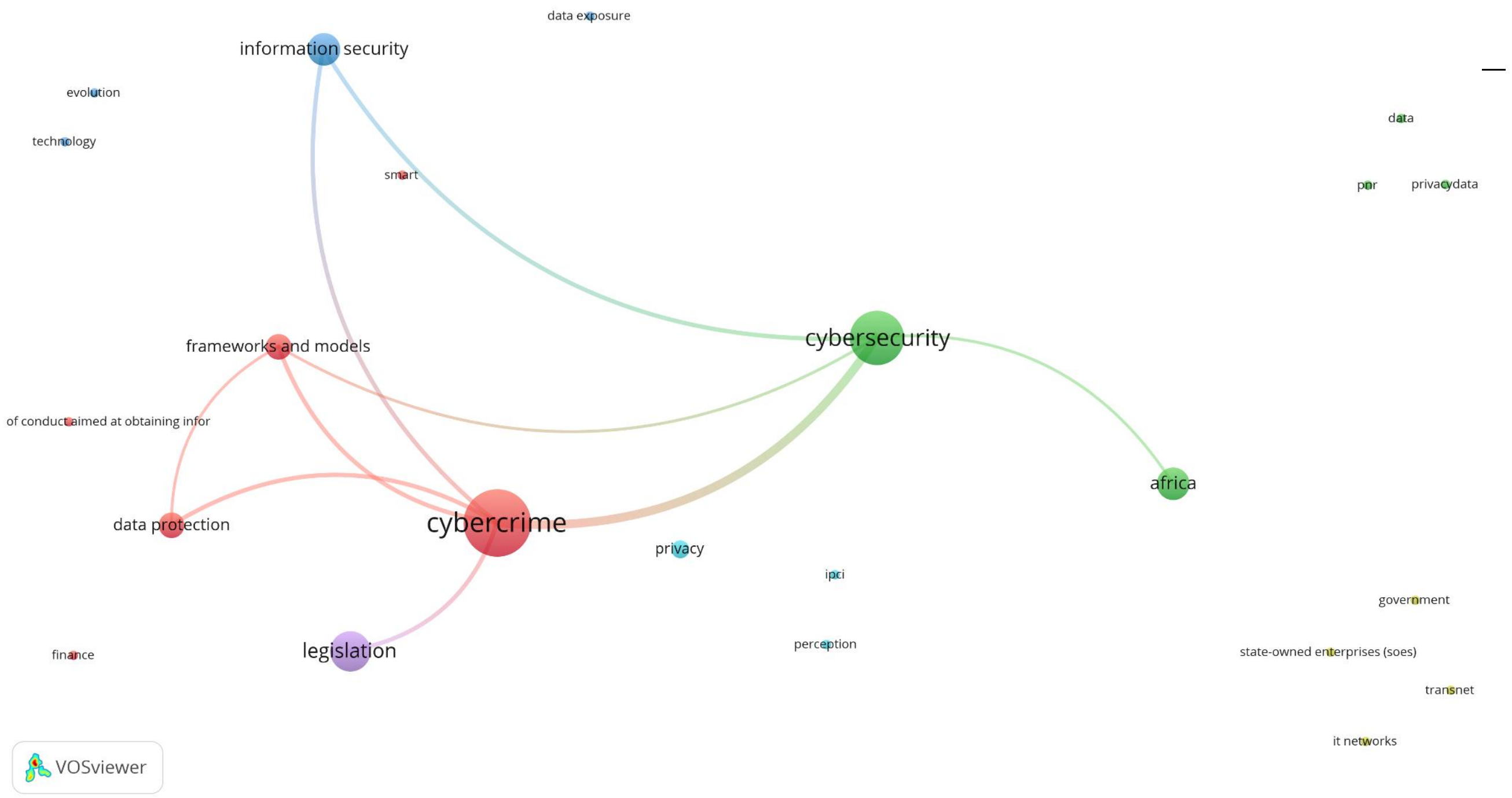

**Figure 7: Network visualisation of cybersecurity state**

Figures 7 shows that "Cybercrime" and "Cybersecurity" were linked to sub-themes such as "Legislation," "Information Security," "Data Protection," and "Frameworks and Models." This supports the notion that the lack of robust cybersecurity legislation contributed to the prevalence of cybercrime in South Africa between 2020 and 2025, particularly in the financial sector (Devanny & Buchan, 2024; Moyana & Chuma, 2023; Mpuru & Kgoale, 2025; Ngwenya & Njenga, 2021; Pieterse, 2021). In contrast, "Government," "Perception," "Privacy," and "Policy" were not linked to other themes, suggesting minimal connections in the collected literature. Additionally, "Trust" was not a prevalent theme in discussions of cybersecurity in South Africa, indicating that while these topics are mentioned, they are not the primary focus of existing research.

## 5. DISCUSSION

This study examined the influence of public trust on the adoption and implementation of cybersecurity, as well as the state of cybersecurity in South Africa. Six main themes were identified: Cybersecurity, Government, Trust, Privacy, Cybercrime, and Public Opinion, each with associated sub-themes and sub-sub-themes, organised according to their relevance to the study and research questions. The results are discussed by addressing each research question, guided by Institutional Trust Theory (Godefroit, Langer & Meuleman, 2017) and the Cyber-Policy Framework (Joubert & Wa Nkongolo, 2025). Institutional Trust Theory considers aspects of citizen trust in institutions, including socialisation indicators, cultural theories, and institutional performance theories; this study focuses on institutional performance, which examines trust in government institutions based on their structures and effectiveness (Godefroit et al., 2017). The Cyber-Policy Framework promotes multi-stakeholder involvement, including public and private sectors and states through a data-centric, cross-cultural approach that can be adopted internationally (Joubert & Wa Nkongolo, 2025). Using the findings alongside these frameworks, the research questions were addressed, and a trust-centric adoption framework was developed. Incorporating trust into cybersecurity policymaking can strengthen the adoption of cybersecurity frameworks in South Africa by addressing weak or indirect linkages identified in the literature.

## Trust-Centric Cybersecurity Adoption Framework (TCCAF)

Figure 8 illustrates the proposed cybersecurity governance model, showing how public opinion is integrated into each stage of the policymaking process.

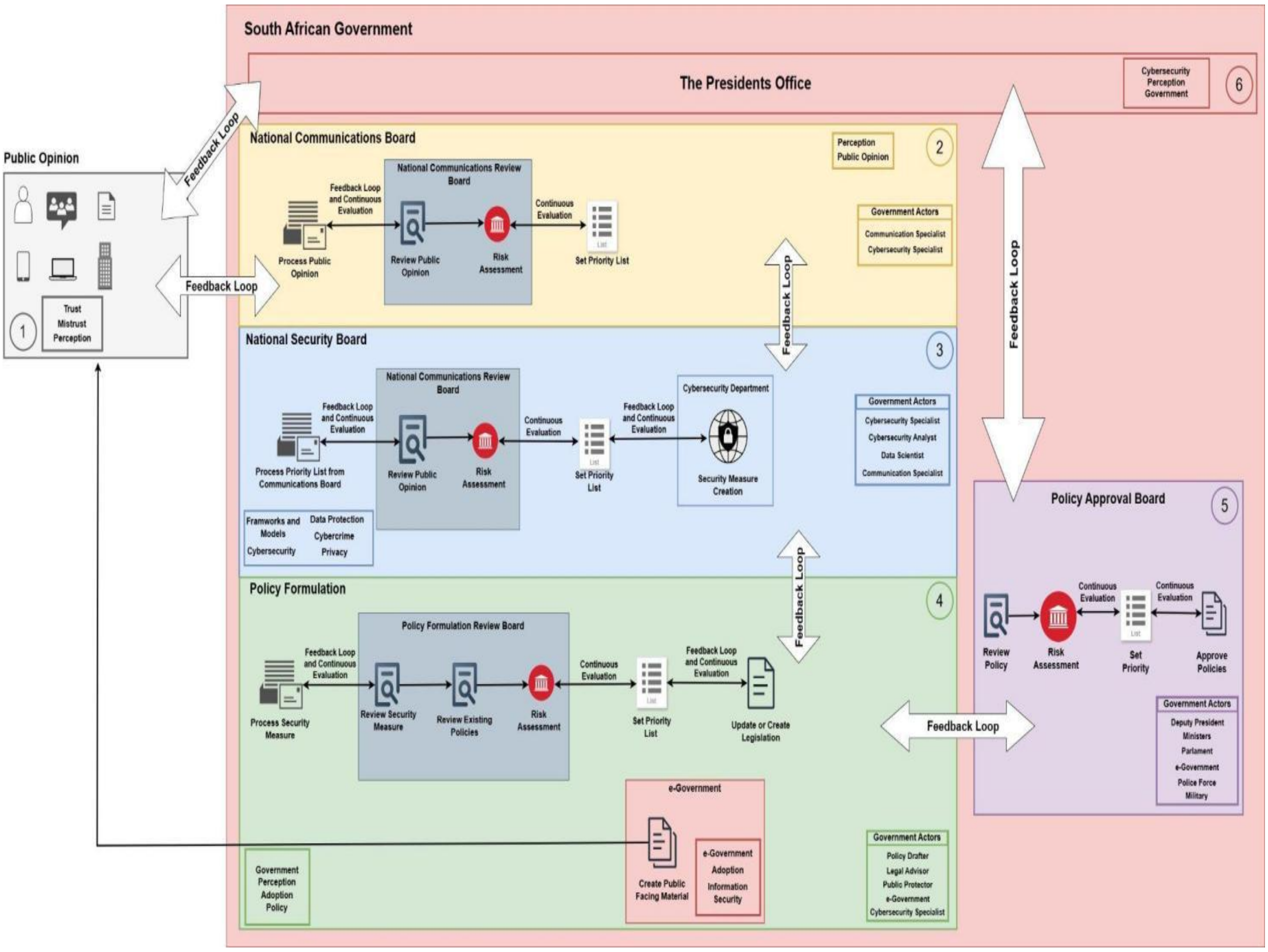

**Figure 8: Trust-centric cybersecurity adoption framework (TCCAF)**

The model consists of interconnected boards including the National Communications Board, the National Security Board, the Policy Formulation Board, and the Policy Approval Board, each with its own internal review mechanisms and continuous feedback loops. Public perception feeds into the system at multiple points, ensuring that trust, concerns, and expectations are reflected in policy evaluations, priority setting, and legislative updates. The e-Government component further strengthens this relationship by providing public-facing information and enabling citizen input. The cyclical flow of feedback both within and between components helps maintain alignment between public trust, policy development, and cybersecurity governance, promoting more legitimate, accountable, and responsive cybersecurity decision-making.

## 6. CONCLUSION

This study examined how public perception specifically trust, mistrust, and distrust affects the adoption and implementation of cybersecurity frameworks in South Africa. A systematic literature review of 34 carefully selected sources revealed that while cybersecurity is a growing national priority, discussions of public trust and citizen perceptions are largely absent in South African scholarship and policy. In contrast, these themes are more prominent in international and broader African contexts, highlighting a gap in local policymaking. Findings suggest that negative perceptions of government institutions can hinder cybersecurity uptake and effectiveness, whereas integrating public trust, technical expertise, and multi-stakeholder participation can strengthen governance and policy legitimacy. Applying Institutional Trust Theory alongside the Cyber-Policy Framework demonstrates that cybersecurity is both a technical and socio-political issue requiring transparency, accountability, and inclusive engagement. The proposed governance model provides practical pathways for embedding public opinion into communication, policy formulation, security review, and approval processes. *The study contributes to the emerging discourse on human-centric cybersecurity in South Africa and lays a foundation for future empirical research, including surveys or participatory approaches to validate and refine the model*. Strengthening trust between citizens and the state is essential for the effective, sustainable, and socially accepted implementation of cybersecurity frameworks.

## 7. RECOMMENDED FUTURE RESEARCH

The study highlights the need to examine how limited cybersecurity awareness and knowledge among citizens affects policy adoption and implementation. The lack of mechanisms for collecting and understanding public opinion suggests exploring the role of social media both as a tool for measuring public perception and for promoting cybersecurity awareness. Addressing these gaps could enhance citizen involvement in cybersecurity policymaking in South Africa and across Africa, while improving the effectiveness and enforcement of existing cybersecurity frameworks.